\documentclass[aps,prl,twocolumn,superscriptaddress]{revtex4-2}

\usepackage{amsthm}
\usepackage{amsmath,bm}
\usepackage{amssymb}
\usepackage{amsfonts}
\usepackage{graphicx}
\usepackage{txfonts}
\usepackage{xcolor}
\usepackage{float}
\usepackage{braket}

\usepackage[colorlinks=true,linkcolor=blue,citecolor=blue,urlcolor=blue]{hyperref}

\newcommand{\ie}{\textit{i.e. }}

\begin{document}

\title{Detection of non-Gaussian quantum correlations through measurement-after-interaction protocols}

\author{Jiajie Guo}
\affiliation{State Key Laboratory for Mesoscopic Physics, School of Physics, Frontiers Science Center for Nano-optoelectronics, $\&$ Collaborative
Innovation Center of Quantum Matter, Peking University, Beijing 100871, China}

\author{Feng-Xiao Sun}
\affiliation{State Key Laboratory for Mesoscopic Physics, School of Physics, Frontiers Science Center for Nano-optoelectronics, $\&$ Collaborative
Innovation Center of Quantum Matter, Peking University, Beijing 100871, China}

\author{Matteo Fadel}
\email{fadelm@phys.ethz.ch}
\affiliation{Department of Physics, ETH Z\"{urich}, 8093 Z\"{urich}, Switzerland}

\author{Qiongyi He}
\email{qiongyihe@pku.edu.cn}
\affiliation{State Key Laboratory for Mesoscopic Physics, School of Physics, Frontiers Science Center for Nano-optoelectronics, $\&$ Collaborative Innovation Center of Quantum Matter, Peking University, Beijing 100871, China}
\affiliation{Collaborative Innovation Center of Extreme Optics, Shanxi University, Taiyuan, Shanxi 030006, China}
\affiliation{Peking University Yangtze Delta Institute of Optoelectronics, Nantong, Jiangsu 226010, China} 
\affiliation{Hefei National Laboratory, Hefei 230088, China}

\begin{abstract}
Additional state evolutions performed before measurement, also called measurement-after-interactions (MAI) protocols, have shown a great potential for increasing the sensitivity of metrological scenarios. Here, we go beyond this result and show that MAI techniques can significantly enhance the detection capability of witnesses for quantum correlations. In particular, we show the possibility of detecting Einstein-Podolsky-Rosen steering and mode entanglement of non-Gaussian states from linear measurements only. Moreover, we show that such approach allows for a significantly higher noise robustness.
\end{abstract}

\maketitle

Experiments with spin ensembles and continuous variable systems commonly perform linear measurements of collective spin components or of phase-space quadratures. 
Compared to measuring higher-order moments, these observables are easier to implement and require significantly fewer statistics to be estimated. 
However, this comes at the cost of insensitivity to non-Gaussian properties of the system's state, which are known to enhance performance in tasks such as quantum computation, communication and sensing~\cite{MariPRL2012,GuoPRA2019,Leenpj2019,StrobelScience2014,XuPRL2022,AndersenNP2015,MattiaPRX2021}.

In this context, one approach to enhance the sensitivity of metrological protocols without requiring the implementation of a complex detection system is to introduce an additional evolution before performing linear measurements \cite{DavisPRL2016,FrowisPRL2016,TommasoPRA2016,NolanPRL2017,MariusQuantum2020,YoucefPRL2021,GuoPRA24}.
The advantages of this so-called Measurement-After-Interaction (MAI), or (Loschmidt) echo, techniques have been successfully demonstrated in a number of experiments \cite{HostenScience2016,BurdScience2019,SimoneNP2022,QiNP2022}.

Inspired by this approach, we want to unlock the potential of MAI techniques for the detection of bipartite quantum correlations that are inaccessible by linear measurements. 
In fact, while for Gaussian states there exist criteria that allow for the detection of entanglement and Einstein-Podolsky-Rosen (EPR) steering from first and second moments of the probability distribution of linear measurement results \cite{ReidPRA1989,Wiseman2007PRL,SimonPRL2000,DuanPRL2000,GiovannettiPRA2003}, criteria that allow for the detection of these correlations in non-Gaussian states require higher order moments, making them much more difficult to derive and to implement \cite{PLamPRL15,HZPRL2006,GuoPRA2023}.

For this reason, rather than looking for criteria that require more sophisticated detection systems, in this work we propose a strategy that consists of deriving criteria that are based on MAI techniques. Concretely, after illustrating the connection between metrological complementarity and the EPR paradox, we show the possibility of detecting EPR steering in non-Gaussian spin states by local measurements of collective spin components when MAI is performed in even only one of the two parties. Then, although we demonstrate that there are continuous-variable cases where MAI does not provide any advantage if the detection is ideal, we show that in the presence of detection noise the MAI strategy significantly improves the noise robustness of the proposed criteria. Furthermore, we demonstrate that these ideas can also be applied to enhance the detection of bipartite mode entanglement.

As we illustrate with concrete and relevant examples, our results can be immediately applied to experiments with both discrete- and continuous-variable systems. This will open the door to the robust detection of entanglement and EPR steering in regimes that are inaccessible by known criteria based on linear measurements and low-order moments of the result's probability distributions. For these reasons, we expect our methods to have a significant impact for the experimental study of quantum correlations in non-Gaussian states of atomic ensembles and electromagnetic fields.

\begin{figure}[t]
    \begin{center}
	\includegraphics[width=90mm]{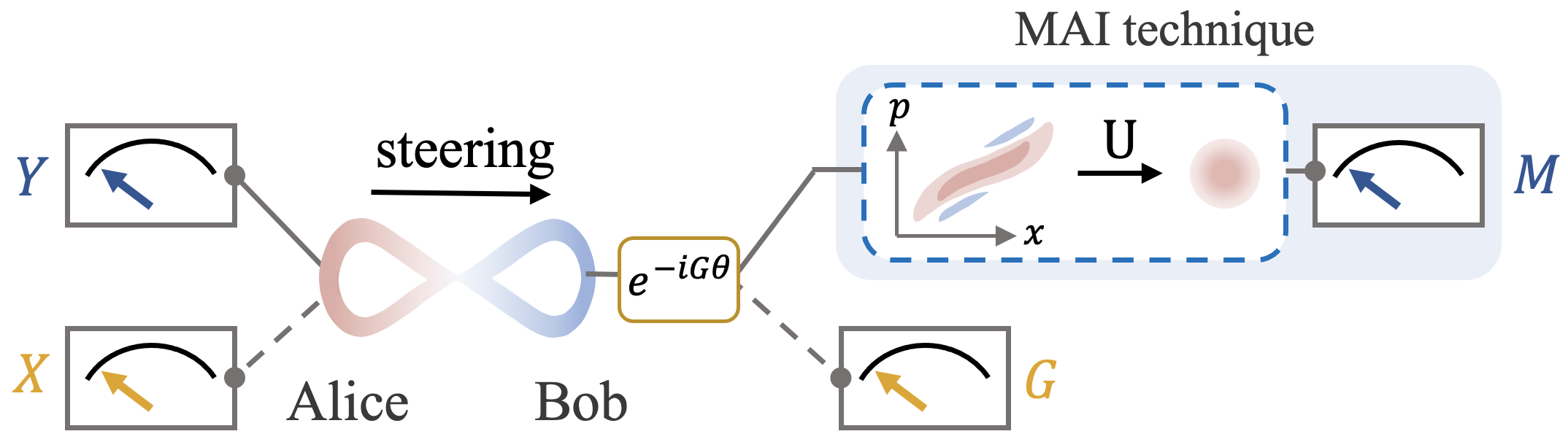}
	\end{center}
    \caption{Illustration a phase estimation protocol assisted by an MAI technique in the framework of EPR paradox. Alice assists Bob in his metrological task by communicating her measurement choice and associated results. Based on this information, Bob will either estimate the generator $G$ or the phase $\theta$ through an MAI technique. In the latter, the probe state will undergo an evolution $U$ before being detected. A violation of complementarity between the estimation uncertainty of $\theta$ and $G$ indicates EPR steering between Alice and Bob.}
    \label{Fig1_Illustration}
\end{figure}

\vspace{2mm}
\textbf{Phase estimation and Measurement-After-Interaction technique.--}
In a typical phase estimation scheme, an unknown phase $\theta$ is encoded into a probe state $\rho$ through the unitary evolution, $\rho(\theta)=e^{-iG\theta} \rho e^{iG\theta}$, where $G$ is the phase-encoding generator. 
A measurement $M$ is then performed on the probe state, and the obtained results are used to infer the phase shift.
Using the method of moments~\cite{PezzeRMP2018}, 
the estimation error after $n$ independent measurements is given by $\text{Var} [\theta_{\text{est}}] =\chi^{2} [\rho,G,M]/n$, where 
\begin{align}\label{eq:SqueezingParameter}
\chi^{2} [\rho,G,M] := \frac{ \text{Var}[\rho,M] }{  | \langle [G,M] \rangle_{\rho}  |^2 } 
\end{align}
is the squeezing parameter. The goal of a metrological protocol is to minimize the uncertainty, typically through the optimization of the state and the measurement. 
For unbiased estimators, the ultimate sensitivity limit for a probe state $\rho$ and phase generator $G$ is given by the quantum Cramér-Rao bound (QCRB) $\text{Var}[\theta_{\text{est}}] \geq \text{Var}[\theta]_{\text{QCRB}} = F_Q[\rho,G]^{-1}$, where $F_Q[\rho,G]$ is quantum Fisher information (QFI)~\cite{BraunsteinPRL1994}. This gives a hierarchy between sensitivities as $\chi^{-2}[\rho,G,M] \leq  F_Q[\rho,G]$.

The squeezing parameter in Eq.~\eqref{eq:SqueezingParameter} is determined by the measurement operator $M$. 
In many experiments this is a linear measurement $M_{\text{L}}$, such as a collective spin operator or a phase-space quadrature. 
These linear measurements $M_{\text{L}}$ have the advantage of being of easy implementation and capable of saturating the QCRB for Gaussian states~\cite{GuoPRA24}.
However, they fail to fully capture the metrological potential of non-Gaussian probe states~\cite{ManuelPRL2019}, where correlations appear in higher-order moments of the distribution.

To overcome this limitation, one possibility is to develop more sophisticated detection systems that implement nonlinear measurements~\cite{ManuelPRL2019}.
Alternatively, another possibility is to perform an MAI protocol where the unitary evolution $U=e^{-iHt}$ is performed on the state right before linear measurement~\cite{DavisPRL2016,FrowisPRL2016,TommasoPRA2016}.
This approach can be understood as the measurement of an MAI operator $M_{\text{MAI}} = U^\dagger M_{\text{L}} U$, whose series expansion contains higher-order moments of the Hamiltonian $H$ and the linear measurement $M_{\text{L}}$ (see Sec. I A in SM~\cite{SM}).
In this way, the MAI squeezing parameter
\begin{align}
\chi_{\text{MAI}}^{2} [\rho,G,M_{\text{MAI}}] := \chi^{2} [\rho,G,M_{\text{MAI}}]
\end{align}
has the potential to surpass its linear counterpart $\chi^{2} [\rho,G,M_{\text{L}}]$, thus allowing to achieve higher sensitivities in phase-estimation protocols.

\vspace{2mm}
\textbf{Metrological task in the framework of EPR paradox.--}
Let us now discuss a phase estimation protocol in the framework of EPR paradox~\cite{MatteoNC2021}. As illustrated in Fig.~\ref{Fig1_Illustration}, Alice and Bob share a quantum state $\rho_{AB}$. 
Alice performs a local measurement $Y$ on her subsystem and obtains result $b$ with probability $p(b|Y)$, which will steer Bob's subsystem in the conditional state $\rho^B_{b|Y}$. 
With information of $Y$ and $b$ given by Alice, Bob can use this conditional state as a probe for a phase-estimation task defined by the local generator $G$ and measurement $M$. 

Defining the assemblage $\mathcal{A}(b,Y)=p(b|Y)\rho^B_{b|Y}$, we introduce the conditional variance \cite{MatteoNC2021}
\begin{align} \label{eq:CondVar}
\text{Var}^{B|A}[\mathcal{A}, M, Y] := \sum_{b} p(b|Y) \text{Var}[\rho^B_{b|Y},M],
\end{align}
which represents the statistical average of the variance for Bob's conditional states if Alice performs a specific measurement $Y$. Analogously to Eq.~\eqref{eq:SqueezingParameter}, we can thus define the average squeezing parameter based on the assemblage $\mathcal{A}$ as 
\begin{align} \label{eq:averagesp}
( \chi^{-2} )^{B|A} [\mathcal{A},  G,M, Y] &:= \frac{ | \langle [ G, M] \rangle_{\rho^B}|^2 }{\text{Var}^{B|A}[\mathcal{A}, M, Y]},
\end{align}
where $\rho^B=\sum_b \mathcal{A}(b|Y)$ is the reduced state of Bob. 

To achieve a higher sensitivity $(\chi^{-2})^{B|A}$, Bob can optimize his local measurement $M$ within a set of accessible operators $\mathbf{M}$. 
Given a family of linear measurements $\mathbf{M}_{\text{L}}$ and a local Hamiltonian $H$, one can always extend it to an MAI family $\mathbf{M}_{\text{MAI}}$ with elements $\{ e^{iHt} \mathbf{M}_{\text{L}}^{(k)} e^{-iHt} \}_k$. 
Optimizing over these families allows us to introduce the maximum average squeezing parameter $( \chi^{-2}_{\text{L}} )^{B|A} := \max_{M\in \text{span} (\mathbf{M}_{\text{L}}) } ( \chi^{-2} )^{B|A} [\mathcal{A},  G,M, Y]$ and $( \chi^{-2}_{\text{MAI}} )^{B|A} := \max_{M\in\text{span} (\mathbf{M}_{\text{MAI}})} ( \chi^{-2} )^{B|A} [\mathcal{A},  G,M, Y]$, respectively.
Since the MAI technique gives access to a larger set of measurement operators, we have the inequality 
\begin{align} \label{eq:sensitivityineq}
( \chi^{-2}_{\text{MAI}} )^{B|A} [\mathcal{A}, G,M_{\text{MAI}}, Y] \geq ( \chi^{-2}_{\text{L}} )^{B|A} [\mathcal{A}, G,M_{\text{L}}, Y] .
\end{align}

The ultimate upper limit on the averaged squeezing parameter is provided by the conditional Fisher information \cite{MatteoNC2021}
\begin{align}
F^{B|A} [\mathcal{A},G,Y] := \sum_{b} p(b|Y) F_Q[\rho^B_{b|Y}, G],
\end{align}
which from the QCRB results in the condition $F^{B|A} [\mathcal{A},G,Y] \geq ( \chi^{-2} )^{B|A} [\mathcal{A}, G,M, Y]$.
Due to the concavity of variance and the convexity of QFI, one can derive $( \chi^{-2} )^{B|A} [\mathcal{A},  G,M, Y] \geq \chi^{-2} [\rho^B, M]$ and $F^{B|A} [\mathcal{A},G,Y] \geq F_Q[\rho^B,G]$, respectively. These inequalities indicate that correlations between the two systems can allow Bob to improve his sensitivity for the local estimation of $\theta$.

\vspace{2mm}
\textbf{Steering criterion with MAI technique.--}
Having established that bipartite quantum correlation can enhance the average metrological sensitivity in one subsystem, in this section we illustrate how the MAI technique can allow for improving the detection capability of EPR steering criteria. 

EPR steering is formalized as the violation of a local hidden state (LHS) model~\cite{ReidRMP2009,Wiseman2007PRL}. Within such a model, an assemblage can be expressed as $\mathcal{A}(a|X)=\sum_{\lambda} p(a|X,\lambda)p(\lambda)\sigma^B_\lambda $, where both Alice's statistics $p(a|X,\lambda)$ and Bob's local state $\sigma^B_\lambda$ depend on a classical variable $\lambda$. 
With the knowledge of Alice's measurement $X$ and results $a$, Bob can construct an estimator to predict the result $g$ for his local measurement $G$. The average estimation error is given as the inference variance $\text{Var}[G_{\text{est}},X]:=\sum_{a,g}p(a,g|X,G)(g_{\text{est}}(a)-g)^2$. Analogously, $\text{Var}[M_{\text{est}},Y]$ can be defined for the other pair of measurements, $Y$ for Alice and $M$ for Bob.

Using the estimator based on the methods of moments, Bob can get the inference variance for the phase after $n$ independent measurements as $\text{Var}[\theta_{\text{est}},Y]:= \frac{\text{Var}[M_{\text{est}},Y]}{n|\langle [G,M]\rangle|^2}$. The uncertainty in simultaneous estimation of phase $\theta$ and generator $G$ is bounded by the complementarity principle, $\text{Var}[\theta_{\text{est}},Y]\text{Var}[G_{\text{est}},X] \geq (4n)^{-1}$. Therefore, one can recover Reid's criterion \cite{ReidPRA1989,ReidRMP2009}
\begin{align}\label{eq:Reidtype}
\Delta^R:=\frac{|\langle[G,M] \rangle_{\rho^B}|^2}{\text{Var}[M_{\text{est}},Y]} -4\text{Var}[G_{\text{est}},X]  \leq 0, 
\end{align}
whose violation reveals EPR steering.
The sharpest formulation of Eq.~\eqref{eq:Reidtype} can be derived from choosing the best estimator. In such case, the inference variance is lower bounded by the conditional variance, e.g. $\text{Var}[G_{\text{est}},X] \geq \text{Var}^{B|A}[\mathcal{A},G,X]$ \cite{ReidRMP2009}.
In typical experiments, linear measurements are used to test Reid's criterion, which combining Eqs.~(\ref{eq:averagesp},\ref{eq:Reidtype}) can be expressed in terms of squeezing parameters
\begin{align}\label{eq:Deltalinear}
\Delta^R_{\text{L}} := ( \chi^{-2}_{\text{L}} )^{B|A} [\mathcal{A}, G,M_{\text{L}}, Y] -4\text{Var}^{B|A} [\mathcal{A}, G, X] \leq 0.
\end{align}
The violation of this inequality provides a sufficient and necessary condition to reveal EPR steering for Gaussian states~\cite{Wiseman2007PRL}. However, since $\Delta^R_{\text{L}}$ is based on only first and second moments of the measurement operators, its applicability to non-Gaussian states is extremely limited.

Keeping in mind the metrological interpretation of the EPR paradox and taking inspiration from the approaches to increase the parameter estimation sensitivity, we now show how MAI techniques can allow for enhancing the detection capabilities of steering criteria.
We consider the scenario where Bob, informed by Alice's $(Y,b)$, will make his local probe state undergo a evolution $U$ before the linear measurements $M$ is performed, see Fig.~\ref{Fig1_Illustration}. 
EPR steering in this MAI-assisted scenario can be detected from the violation of 
\begin{align} \label{eq:DeltaMAI}
\Delta^R_{\text{MAI}} := ( \chi^{-2}_\text{MAI} )^{B|A} [\mathcal{A}, G,M_\text{MAI}, Y] -4\text{Var}^{B|A} [\mathcal{A}, G, X] \leq 0.
\end{align}
Crucially, although this criterion does not require any change in the detection system compared to $\Delta^R_{\text{L}}$, we will show that the additional evolution preceding the measurement allows for the detection of EPR steering in significantly non-Gaussian states.

Here, it is worth mentioning that the ultimate limit on the phase estimation error is given by the conditional Fisher information, i.e. $n\text{Var}[\theta_{\text{est}},Y ]\geq (F^{B|A}[\mathcal{A},G,Y])^{-1}$. 
This relation leads to the EPR steering criterion~\cite{MatteoNC2021}
\begin{equation}\label{eq:DeltaF}
    \Delta^F:=F^{B|A}[\mathcal{A},G,Y]-4\text{Var}^{B|A} [\mathcal{A}, G, X] \leq 0 .
\end{equation}
Although this criterion is generally more powerful than Reid's criterion in Eq.~\eqref{eq:Deltalinear} and the MAI criterion in Eq.~\eqref{eq:DeltaMAI}, it requires measuring the quantum Fisher information. Since this is equivalent to performing full state tomography, a notoriously challenging task in many experiments, directly testing criterion \eqref{eq:DeltaF} may be practically infeasible.

\vspace{2mm}
\textbf{Measurement optimization.--}
Given a family of experimentally accessible operators $\mathbf{G}=\left(G_1,\cdots,G_K \right)$, we express the measurement $G$ as a linear combination of these operators $G=\mathbf{n}^T \mathbf{G}$, with $\mathbf{n}$ a normalized real vector. 
Analogously, we express $M=\mathbf{m}^T \mathbf{M}$. 
Optimizing over $\mathbf{m}$, the average sensitivity Eq.~\eqref{eq:averagesp} reads 
\begin{align}
\max_{M \in \text{span} (\mathbf{M}) } (\chi^{-2})^{B|A}[ \mathcal{A},G,M , Y] = \mathbf{n}^T \mathcal{M} [\mathcal{A}, \mathbf{G}, \mathbf{M},Y ] \mathbf{n} ,
\end{align}
where we have used the moment matrix  $\mathcal{M} [\mathcal{A}, \mathbf{G}, \mathbf{M},Y ] := \mathbf{C} [\rho^B,\mathbf{G},\mathbf{M}]^T \mathbf{\Gamma}^{B|A} [\mathcal{A}, \mathbf{M}, Y]^{-1} \mathbf{C} [\rho^B,\mathbf{G},\mathbf{M}]$ \cite{ManuelPRL2019}. Here, the commutator matrix has elements $\mathbf{C} [\rho^B,\mathbf{G},\mathbf{M}]_{ij}=-i\langle [M_j,G_i]\rangle_{\rho^B}$, and the conditional variance matrix has elements $\mathbf{\Gamma}^{B|A} [\mathcal{A}, \mathbf{M}, Y]_{ij}:=\sum_{b}p(b|Y) \text{Cov}(M_i,M_j)_{\rho^B_{b|Y}}$. The conditional variance in the second term of Eq.~\eqref{eq:DeltaMAI} can then also be written as $\text{Var}[\mathcal{A},G,X]=\mathbf{n}^T \mathbf{\Gamma}^{B|A} [\mathcal{A}, \mathbf{G}, X] \mathbf{n}$. In this way, Reid's criterion can be expressed in terms of matrices and vectors associated with accessible operators, and thus in a convenient form for being systematically optimized  (see Sec. I A in SM~\cite{SM}).

Let us consider now the case of an MAI-assisted protocol. We first introduce a complete family of linear observables $\mathcal{L}$ and, for simplicity, we take $\mathbf{G}\subseteq \mathcal {L}$ (i.e. linear-encoding) and  $X,Y \in \text{span}(\mathcal{L})$. A time-dependent MAI family $\mathbf{M}_{\text{MAI}}(t)$ contains  elements $\mathbf{M}_{\text{MAI}}^{(k)}(t):=e^{iHt} \mathcal{L}^{(k)} e^{-iHt}$, with a given local Hamiltonian $H$. In this case, we can derive the maximum violation $\delta^R_{\text{MAI}}$ of Eq.~\eqref{eq:DeltaMAI} as
\begin{align}
\max_{X,Y\in \text{span}(\mathcal{L}), t} \lambda_{\max} \big( \mathcal{M}[\mathcal{A}, \mathbf{G},\mathbf{M}_{\text{MAI}}(t),Y]-4\mathbf{\Gamma}^{B|A}[\mathcal{A},\mathbf{G},X]   \big).
\end{align}
The maximum violation $\delta^R_{\text{L}}$ of Eq.~\eqref{eq:Deltalinear} can be derived analogously by setting $t=0$. Furthermore, as a benchmark value, we consider the quantity $\delta^F:=\max \Delta^F$, whose derivation is in Sec. II of the SM~\cite{SM}.

\begin{figure}[t]
    \begin{center}
	\includegraphics[width=85mm]{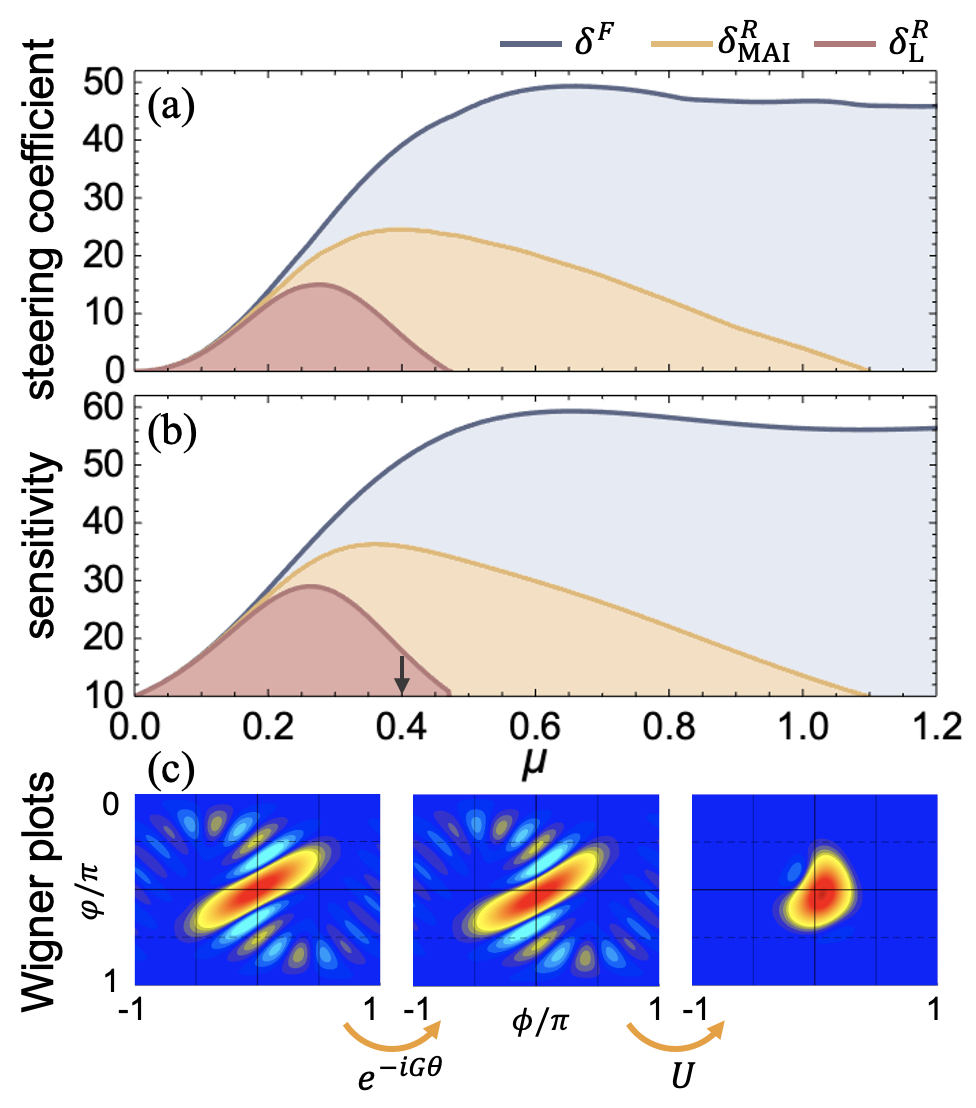}
	\end{center}
    \caption{Comparison between steering criteria and average sensitivities as a function of $\mu$, for a split spin squeezed states with $N=20$. \textbf{(a)} maximum violation of the steering criteria showing the hierarchy $\delta^R_{\text{L}}\leq \delta^R_{\text{MAI}} \leq \delta^F$. \textbf{(b)} first term in the criteria, showing the average sensitivities $ (\chi^{-2}_{\text{L}})^{B|A}[\mathcal{A},G,M_{\text{L}},Y] \leq (\chi^{-2}_{\text{MAI}})^{B|A}[\mathcal{A},G,M_{\text{MAI}},Y] \leq F^{B|A}[\mathcal{A},G,Y]$. \textbf{(c)} for $\mu=0.4$, Wigner functions after encoding and MAI interaction using Bob's conditional states $\rho^B_{(l_A=5,N_A=10|Y)}$.}
    \label{Fig2_splitspinstate}
\end{figure}

\vspace{2mm}
\textbf{Steering of atomic split spin squeezed states.--}
We benchmark our method with the paradigmatic case of atomic split spin squeezed states, which have recently attracted increasing interest, both theoretically and experimentally \cite{YumangNJP2019,MatteoScience2018,PaoloPRX2023,GuocatPRA2023}.

An ensemble of $N$ spin-1/2 particles is initially prepared in a coherent spin state polarized along the $x$ direction, and then evolve via the one-axis twisting (OAT) Hamiltonian $H_{\text{OAT}}=\hbar \chi S_z^2$, with collective spin operator $S_z= \frac{1}{2}\sum_{i=1}^N \sigma^{(i)}_z$. Finally, the ensemble is spatially split by a beam-splitter-like transformation. The resulting split spin squeezed state reads \cite{YumangNJP2019}
\begin{align}\label{eq:SSS}
\ket{\Phi(\mu)} = \frac 1 {2^{N}} \sum_{N_A=0}^{N}&\sum_{k_A=0}^{N_A}\sum_{k_B=0}^{N-N_A}  \sqrt{\binom{N}{N_A} \binom{N_A}{k_A} \binom{N-N_A}{k_B}} \notag\\ &\quad\times e^{- i \frac{\mu}{2} (N/2 - k_A - k_B)^2} \ket{k_A}_{N_A} \ket{k_B}_{N-N_A},
\end{align}
where $N_{\alpha}$ is the number of atoms in $\alpha=\{A,B\}$ side, and the total number is conserved as $N=N_A+N_B$. $|k_{\alpha}\rangle_{N_\alpha}$ is local Dicke state with $k_{\alpha}$ excitations. $\mu=2\chi t $ characterizes the squeezing generated by OAT evolution. As $\mu$ increases, the state will be squeezed from Gaussian to highly non-Gaussian states~\cite{YumangNJP2019,guo_detecting_2023}. 

A complete family of linear observables for atomic ensembles consists of all incompatible collective spin components, namely the set $\mathcal{L}=\left(S_x, S_y, S_z \right)$. We will construct our measurement operators from linear combinations of these observables. 
For the MAI technique, we introduce a second OAT evolution with opposite twisting direction, $U_{\text{MAI}}=e^{i  \frac{\mu_2}{2} \left( S_z^B\right)^2 }$, where the MAI evolution time $\mu_2$ can be optimized. We have shown in Sec. I B in SM~\cite{SM} that such OAT interaction along $z$-axis is an efficient operator within accessible unitary family.
In Fig.~\ref{Fig2_splitspinstate}(a) we show the different steering criteria as a function of OAT squeezing $\mu$, where positive values indicate the detection of EPR steering. 
We observe that all criteria perform well for Gaussian spin states, \ie $\mu \lesssim 0.2$. However, in the non-Gaussian region $\mu\gtrsim 0.2$, linear Reid's criterion $\delta^R_{\text{L}}$ will soon stop being useful. 
On the other hand, the criterion assisted by MAI-technique $\delta^R_{\text{MAI}}$ shows the capability of detecting steering over a significantly wider class of non-Gaussian states. 

Although $\delta^{F}$ outperforms both previous criteria, $\delta^R_{\text{MAI}}$ is significantly more experimentally practical because it only requires the same linear measurements as Reid's criterion. 
The average sensitivity coming from the first term in the criteria is presented in Fig.~\ref{Fig2_splitspinstate}(b), showing the main reason behind the observed hierarchy.
As an illustration, we show in Fig.~\ref{Fig2_splitspinstate}(c) the Wigner plot of one of Bob's conditional states during the MAI protocol. 
The OAT evolution in MAI scheme transforms the state from a highly over-squeezed non-Gaussian state to a coherent-state-like state while amplifying the phase signal $\theta$. 

In realistic experiments, the quantum states interact with the environment, so that the MAI evolution can not be characterized by the unitary operations. The MAI protocol under general quantum channels is discussed in Sec. I C of SM~\cite{SM}. As a concrete example, we show their robustness against the atom loss, a common noise channel in atomic experiments.

\begin{figure}[t]
    \begin{center}
	\includegraphics[width=85mm]{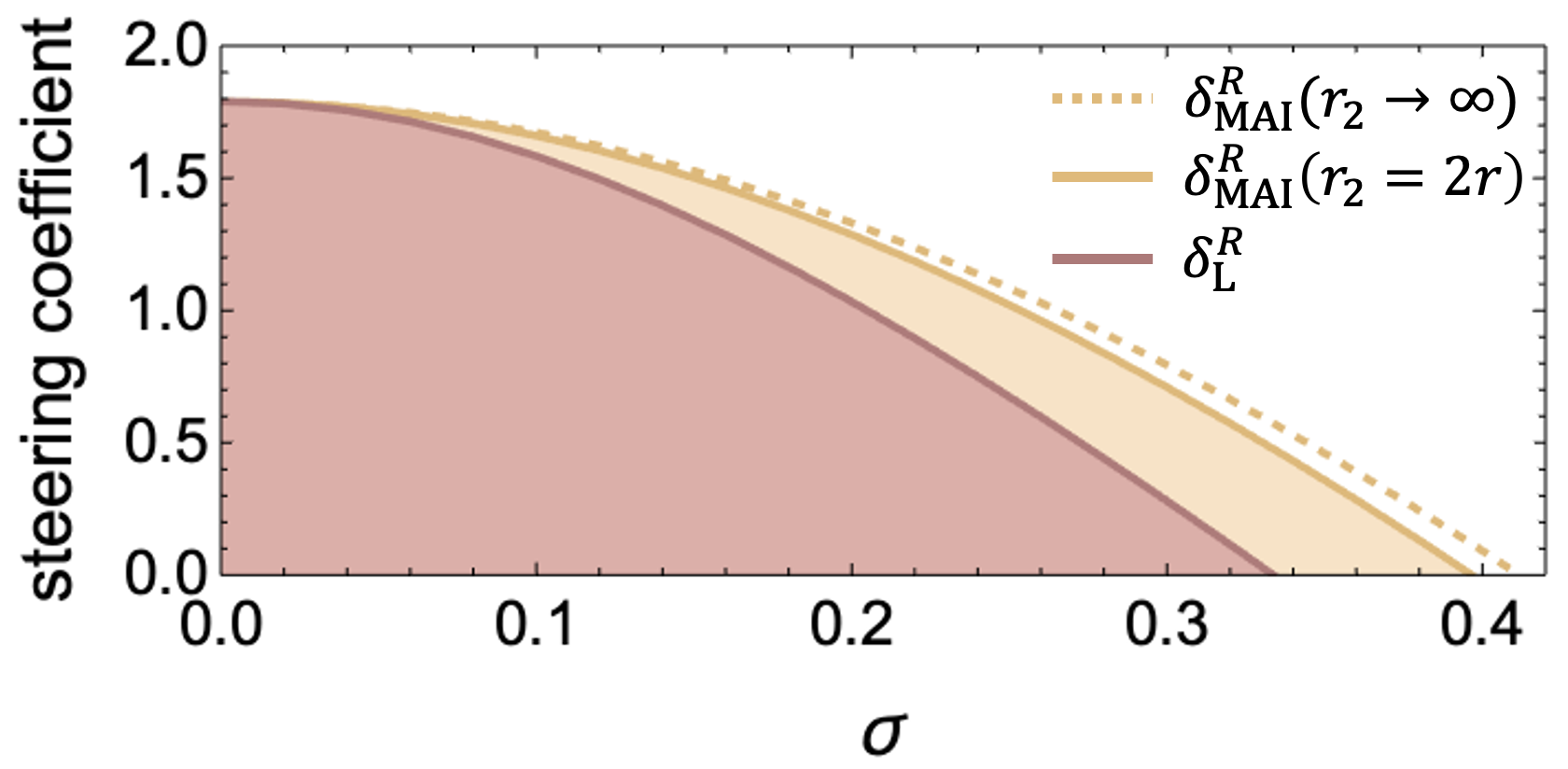}
	\end{center}
    \caption{Robustness to detection noise. Maximum violation of the steering criteria $\delta^R_{\text{MAI}},\delta^R_{\text{L}}$ as a function of the standard derivation of detection noise $\sigma$ for squeezing $r=0.5$. Noise robustness increases as the squeezing $r_2$ in MAI technique increases.}
    \label{Fig3_CVsqueezedvacuum}
\end{figure}

\vspace{2mm}
\textbf{Steering of two-mode squeezed states.--}
The usefulness of our method can be illustrated in continuous variable (CV) cases as well.
We take as an example the two-mode squeezed (TMS) vacuum state $|\psi\rangle=S_{AB} |0,0\rangle$ prepared by applying the TMS operator $S_{AB}=\exp\left[-\xi a^\dagger b^\dagger+ \xi^* a b \right]$ on the vacuum, where $\xi=r e^{i\theta}$ is the squeezing parameter. 
For Gaussian states a linear estimator is optimal \cite{ReidRMP2009}, meaning that the conditional variance Eq.~\eqref{eq:CondVar} becomes $\text{Var}^{B|A} [\mathcal{A},M,Y]= \text{Var} [M+g_Y Y]$, where $g_Y$ is a real constant that is chosen to minimize the conditional variance. Therefore, Reid's criterion can be expressed in a linear-estimation form as 
\begin{align}\label{eq:linearReid}
\Delta^R=\frac{|\langle [G,M ]\rangle_{\rho^B}|^2}{\text{Var}[M+g_Y Y] } -4\text{Var}[G+g_X X] \leq 0.
\end{align}

To characterize a CV state, a complete family of linear quadrature observables is $\mathcal{L}=\{ x,p \}$. 
The evolution in the MAI protocol is taken to be analogous to the one used for the state preparation, namely we consider a single-mode squeezing operation $S_B=\exp\left[-\xi_2 (b^{\dagger})^2+\xi_2^* b^2 \right]$, with $\xi_2=r_2 e^{i\theta_2}$. 
After optimizing the parameters $g_X,g_Y$, we find for the steering coefficients $\delta^R_{\text{L}}=\delta^R_{\text{MAI}}=2\left(\cosh(2r)-1/\cosh(2r)\right)$, meaning that both schemes will detect steering if $r>0$. 
Note that this result is independent of the squeezing $r_2$, showing that using the MAI strategy does not result in a larger violation. 
This is consistent with what is known in the context of single-mode metrology \cite{GuoPRA24,fadelCV24}. 
However, if we look at a realistic scenario where detection noise is present, we can show that the MAI-assisted protocol gives a clear advantage. 
Consider Gaussian detection noise, which leads to fluctuations with standard deviation $\sigma$ on the outcomes. 
Under the influence of this noise, the first term in Eq.~\eqref{eq:linearReid} becomes 
\begin{align}\label{eq:firstlinearReid}
\max_{g_Y,Y,G} \frac{|\langle [G,M_{\text{MAI}} ]\rangle_{\rho^B}|^2}{\text{Var}[M_{\text{MAI}}+g_Y Y] }=\frac{e^{2r_2}}{\frac{e^{2r_2}}{2\cosh{(2r)}} + (1+e^{2r_2} \tanh^2{(2r)})\sigma^2},
\end{align}
which will recovers the result without MAI when $r_2=0$. 
The average sensitivity in Eq.~\eqref{eq:firstlinearReid} causes the violation of $\delta^R_{\text{MAI}}$ to increase with $r_2$. As $r_2$ approaches infinity, $\delta^R_{\text{MAI}}$ reaches its maximum.
In Fig.~\eqref{Fig3_CVsqueezedvacuum}, we compare the maximum violation of the two criteria as a function of standard derivation of detection noise $\sigma$. Assisted by MAI technique, $\delta^R_{\text{MAI}}$ significantly outperforms the typical linear criterion $\delta^R_{\text{L}}$ in terms of detection noise robustness (see Sec. I D in SM~\cite{SM}).

\begin{figure}[t]
    \begin{center}
	\includegraphics[width=85mm]{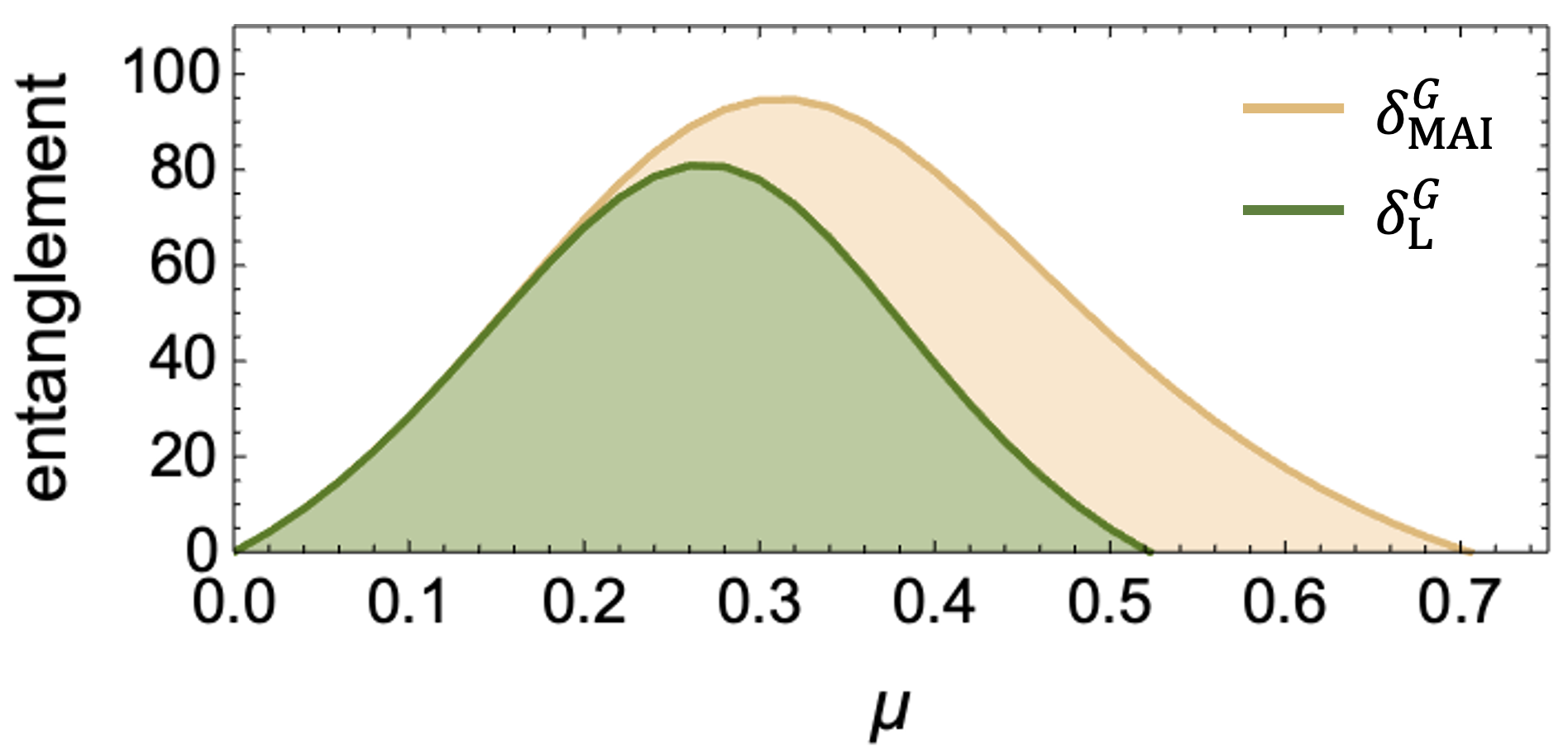}
	\end{center}
    \caption{Mode entanglement in a split spin squeezed state with $N=20$ atoms. Comparison of the violation of Giovannetti's criterion under the typical and MAI protocols as a function of $\mu$.  }
    \label{Fig4_entanglement}
\end{figure}

\vspace{2mm}
\textbf{Detect mode entanglement with local MAI.--}
Besides enhancing the capability of steering criteria, we show that the MAI-assisted scheme is also useful for improving mode-entanglement detection. Compared to EPR steering, entanglement is a weaker type of quantum correlation that it is easier to prepare and detect experimentally. We detect mode entanglement for split spin squeezed states in Eq.~\eqref{eq:SSS} by Giovannetti's criterion \cite{YumangNJP2019,fadelPRA20,MatteoScience2018}, which can be written as 
\begin{align}
\Delta^G := \frac{ \big||g_X g_Y| \langle [X,Y]\rangle+\langle [G,M]\rangle  \big|^2}{ \text{Var}[M+g_Y Y] } -4\text{Var}[G+g_X X] \leq 0.
\end{align}
Analogously, we use notations $\delta^G_{\text{L}},\delta^G_{\text{MAI}}$ for the maximum violation using linear observables and the MAI technique, respectively (see Sec. III in SM~\cite{SM}). 
The relationship $\delta^G_{\text{MAI}} \geq \delta^G_{\text{L}}$ is evident form Fig.~\ref{Fig4_entanglement}. This implies a significant advantage provided by $\delta^G_{\text{MAI}}$ for the detection of non-Gaussian quantum correlations.

\vspace{2mm}
\textbf{Conclusions.--} 
We have introduced a strategy to robustly detect bipartite entanglement and EPR steering in non-Gaussian states, which is based on implementing an MAI protocol on one of the two parties. 
Crucially, this further evolution does not involve any additional assumptions, since entanglement can not increase under local operations and classical communication.
By analyzing experimentally relevant examples of discrete- and continuous-variable systems, we demonstrate that our strategy shows improved ability to detect non-Gaussian correlations and increased robustness to detection noise. Crucially, this is achieved without any change in the detection system or in the amount of data to be collected.

Our method finds immediate applications in experiments with atomic ensembles and electromagnetic fields, where MAI techniques have already been explored for metrology.
Potentially, this approach could also be applied to the detection of Bell nonlocality \cite{StobinskaPRA07,PaternostroPRA09,KitzingerPRA21} or multipartite steering \cite{Armstrong15,TehPRA22}.
Our work provides a powerful tool for exploring quantum correlations in non-Gaussian states and for unlocking their use in quantum technologies.

\vspace{2mm}
\textit{Acknowledgments.--} 
This work was supported by the National Natural Science Foundation of China (No. 12125402, No. 12474256, No. 12447157), Beijing Natural Science Foundation (Grant No. Z240007), and the Innovation Program for Quantum Science and Technology (No. 2021ZD0301500). J.G. acknowledges Postdoctoral Fellowship Program of CPSF (GZB20240027), and the China Postdoctoral Science Foundation (No. 2024M760072). M.F. was supported by the Swiss National Science Foundation Ambizione Grant No. 208886, and by The Branco Weiss Fellowship -- Society in Science, administered by the ETH Z\"{u}rich.

\bibliographystyle{apsrev4-1} 
\bibliography{Reference}

\clearpage
\newpage

\begin{widetext}

\section{Supplemental material for ``Detection of non-Gaussian quantum correlations through measurement-after-interaction protocols''}

\section{I.\quad Reid's criterion}
In the framework of EPR paradox, informed with a specific measurement $X$ and corresponding result $a$ from Alice, Bob can choose a local measurement $G$ on his side, and make an estimator $g_{\text{est}}(a)$ for the result $g$. The resulting average deviation is called inference variance $\text{Var}[G_{\text{est}},X] := \sum_{a,g} p(a,g|X,G) (g_{\text{est}}(a) -g)^2$, where $p(a,g|X,G)$ is the joint probability distribution.  Analogously, We can also define $\text{Var}[M_{\text{est}},Y]$. To describe the complementarity between two local observables $M$ and $G$, Reid's criterion can be expressed as
\begin{align}
\Delta^R:= \frac{ |\langle [G,M] \rangle_{\rho^B}|^2 }{\text{Var} [M_{\text{est}},Y] } -4\text{Var}[G_{\text{est}},X] \leq 0 
\end{align}

Furthermore, if one chooses the optimal estimator constructed with Bob's conditional states $\rho_{a|X}^B$ as $g_{\text{est}}(a)=\text{Tr}[\rho_{a|X}^B G]$, the inference variance $\text{Var}[G_{\text{est}},X]$ can be minimized. Under such estimator, one can define the conditional variance
\begin{align}
\text{Var}^{B|A} [ \mathcal{A}, G, X ]:= \sum_a p(a|X) \text{Var}[\rho^B_{a|X},G], 
\end{align}
which provides a lower bound to the inference variance $\text{Var}[G_{\text{est}},X] \geq \text{Var}^{B|A} [ \mathcal{A}, G, X ]$. This can yield a sharper formulation of Reid's criterion
\begin{align} \label{eq:Reidsharper}
( \chi^{-2} )^{B|A} [\mathcal{A},G,M,Y] -4\text{Var}^{B|A}[\mathcal{A},G,X]  \leq 0,
\end{align}
where
\begin{align}
( \chi^{-2} )^{B|A} [\mathcal{A},G,M,Y] := \frac{ |\langle [G,M] \rangle_{\rho^B}|^2 }{ \text{Var}^{B|A} [\mathcal{A},M,Y]}
\end{align}
is the average squeezing parameter quantifying the average sensitivity for the assemblage $\mathcal{A}$.

\subsection{A. Measurement optimization for Reid's criterion}
Let us consider a family of accessible operators $\mathbf{G}=\{G_1,\cdots,G_K\}$, such that any operator within the setting can be expressed by a linear combination as $G=\mathbf{n}^T \mathbf{G}=\sum_{k=1}^K n_k G_k$. Similarly, we can also write $M=\mathbf{m}^T\mathbf{M}$. In this way, the average squeezing parameter, i.e. the first term of Reid's criterion in Eq.~\eqref{eq:Reidsharper}, can be written as
\begin{align}
( \chi^{-2} )^{B|A} [\mathcal{A},G,M,Y] &:= \frac{ |\langle [G,M] \rangle_{\rho^B}|^2 }{ \text{Var}^{B|A} [\mathcal{A},M,Y]} \nonumber \\
&= \frac{ |\mathbf{n}^T \mathbf{C}[\rho^B, \mathbf{G},\mathbf{M}] \mathbf{m} |^2 }{ \sum_b p(b|Y) \mathbf{m}^T \boldsymbol{\Gamma}[\rho^B_{b|Y},\mathbf{M}] \mathbf{m} } \nonumber \\
&= \frac{ | \mathbf{n}^T \mathbf{C}[\rho^B, \mathbf{G},\mathbf{M}] \mathbf{m} |^2 }{ \mathbf{m}^T \boldsymbol{\Gamma}^{B|A}[\mathcal{A},\mathbf{M},Y] \mathbf{m} }.
\end{align}
Here $\mathbf{C}[\rho^B, \mathbf{G},\mathbf{M}]$ is the commutator matrix for Bob's reduced states $\rho^B$, which contains the elements $\mathbf{C}[\rho^B, \mathbf{G},\mathbf{M}]_{ij}=-i \langle [ M_j,G_i ] \rangle_{\rho^B}$.  $\boldsymbol{\Gamma}^{B|A}[\mathcal{A},\mathbf{M},Y] = \sum_b p(b|Y) \boldsymbol{\Gamma}[\rho^B_{b|Y},\mathbf{G}]$ is the covariance matrix for the assemblage $\mathcal{A}$, where $\boldsymbol{\Gamma}[\rho^B_{b|Y},\mathbf{G}]_{ij}=\text{Cov}(G_i,G_j)_{\rho^B_{b|Y}}$. Using the method proposed in~\cite{ManuelPRL2019}, the average squeezing parameter under the optimization over $M$ can be solved as
\begin{align} \label{eq:Reid1stterm}
\max_{M\in \text{span}(\mathbf{M})} ( \chi^{-2} )^{B|A} [\mathcal{A},G,M,Y] = \mathbf{n}^T \mathcal{M} [\mathcal{A},\mathbf{G},\mathbf{M},Y] \mathbf{n},
\end{align}
where we use the moment matrix
\begin{align}
\mathcal{M} [\mathcal{A},\mathbf{G},\mathbf{M},Y] = \mathbf{C}[\rho^B,\mathbf{G},\mathbf{M}^T] \boldsymbol{\Gamma}^{B|A}[\mathcal{A},\mathbf{M},Y]^{-1} \mathbf{C}[\rho^B,\mathbf{G},\mathbf{M}].
\end{align}
The conditional variance, the second term in Reid's criterion in Eq.~\eqref{eq:Reidsharper}, can be expressed by covariance matrix
\begin{align}\label{eq:Reid2ndterm}
\text{Var}^{B|A} [\mathcal{A},G,X] = \mathbf{n}^T \boldsymbol{\Gamma}^{B|A} [\mathcal{A},\mathbf{G},X] \mathbf{n}.
\end{align}
Combining Eq.~\eqref{eq:Reid1stterm} and Eq.~\eqref{eq:Reid2ndterm}, we can optimize Reid's criterion over the phase generator $G$, which reads
\begin{align}
\max_{M\in \text{span}(\mathbf{M}), G\in \text{span}(\mathbf{G})} \Delta^{R} &=  \max_{\mathbf{n}} \left( \mathbf{n}^T \mathcal{M} [\mathcal{A},\mathbf{G},\mathbf{M},Y] \mathbf{n} - 4\mathbf{n}^T \boldsymbol{\Gamma}^{B|A} [\mathcal{A},\mathbf{G},X] \mathbf{n} \right) \nonumber \\
&= \lambda_{\max} \left( \mathcal{M} [\mathcal{A},\mathbf{G},\mathbf{M},Y] - 4\boldsymbol{\Gamma}^{B|A} [\mathcal{A},\mathbf{G},X]  \right)
\end{align}
After optimizing Alice's measurement $X,Y$ within a family of linear observables $\mathcal{L}$, we can finally obtain the maximum violation for Reid's criterion
\begin{align}\label{eq:deltaRmax}
\delta^R := \max_{X,Y\in\text{span}(\mathcal{L})} \lambda_{\max} \left( \mathcal{M} [\mathcal{A},\mathbf{G},\mathbf{M},Y] - 4\boldsymbol{\Gamma}^{B|A} [\mathcal{A},\mathbf{G},X]  \right)
\end{align}
Here, the eigenvector corresponding to the maximum eigenvalue in Eq.~\eqref{eq:deltaRmax} gives the optimal direction $\mathbf{n}_{\text{opt}}$ for Bob's measurement $G$. The optimal direction for the other measurement $M$ is given as $\mathbf{m}_{\text{opt}}=\gamma \boldsymbol{\Gamma}^{B|A}[\mathcal{A},\mathbf{M},Y]^{-1}\mathbf{C}[\rho^B,\mathbf{G},\mathbf{M}]\mathbf{n}_{\text{opt}} $ with a real normalization value $\gamma$.

In the atomic ensemble we considered in the main text, a complete family of linear observables is $\mathcal{L}=\{S_x,S_y,S_z\}$, with collective spin observables $S_{\alpha}=\frac{1}{2}\sum_{i} \sigma_{\alpha}^{(i)} $ on a sub-ensemble. In a typical estimation scenario where only linear observables are considered, we set both of Bob's measurements settings as $\mathbf{G}=\mathcal{L}$ and $ \mathbf{M}_\text{L}=\mathcal{L}$. The squeezing of the probe lies in $yz-$plane, to simplify the calculation, we can restrict Alice's measurements to this plane. In this case, Alice's measurements can be determined by  $X(\theta_X)=\cos{(\theta_X)}S_y+\sin{(\theta_X)}S_z$ and $Y(\theta_Y)=\cos{(\theta_Y)}S_y+\sin{(\theta_Y)}S_z$, where $\theta_{X,Y}\in[0,\pi]$. Therefore, the maximum violation for Reid's criterion with linear operators involved is
\begin{align}\label{eq:deltaRLmax}
\delta^R_{\text{L}} = \max_{\theta_X,\theta_Y} \lambda_{\max} \left( \mathcal{M} [\mathcal{A},\mathbf{G},\mathbf{M}_{\text{L}},Y] - 4\boldsymbol{\Gamma}^{B|A} [\mathcal{A},\mathbf{G},X]  \right).
\end{align}

While in the scenario with the MAI technique, we introduce an additional unitary evolution $U(t)=e^{-iHt}$ before linear measurement, so that the process can be described by an MAI operator $M_{\text{MAI}} = U^\dagger M_{\text{L}} U $. Mathematically, this MAI operator can be further expanded as
\begin{align}
M_{\text{MAI}} &= M_{\text{L}} +it[H, M_{\text{L}}] - \frac{t^2}{2} [H,[H,M_{\text{L}}]] + \mathcal{O}(t^3).
\end{align}
With these higher-order moments involved, MAI detection is more sensitive to higher-order correlations, thereby enhancing the detection capability of MAI-based criteria.

We can construct a family of MAI operators $\mathbf{M}_{\text{MAI}}(t)=\{ U^\dagger(t) S_x U(t), U^\dagger(t) S_y U(t), U^\dagger (t) S_z U(t)  \} $. The family $\mathbf{M}_{\text{MAI}}(t)$ is time-dependent, and will return to $\mathbf{M}_{\text{L}}$ if $t=0$. Thus, the maximum violation of Reid's criterion with MAI assistance can be derived as
\begin{align}
\delta^R_{\text{MAI}} = \max_{\theta_X,\theta_Y,t} \lambda_{\max} \left( \mathcal{M} [\mathcal{A},\mathbf{G},\mathbf{M}_{\text{MAI}}(t),Y] - 4\boldsymbol{\Gamma}^{B|A} [\mathcal{A},\mathbf{G},X]  \right).
\end{align}
Since MAI technique considers the optimization over times $t$, an inequality relation is given as $\delta^R_{\text{MAI}} \geq \delta^R_{\text{L}}$.

\subsection{B. Optimization over the second OAT interaction }

In the main text, we have investigated the simplest non-trivial unitary, namely a local OAT interaction along the fixed $z-$axis. 
A natural next step would be to consider if other unitaries could further improve the detection sensitivity of the proposed protocol. 
However, finding the optimal unitary operation is challenging, and may lead to operations that are not experimentally feasible. 
For this reason, we decide to perform such an optimization while restricting it to a family of accessible unitary operations. 

In atomic systems using collective spin, commonly available unitary operations are collective rotations and squeezing (such as OAT interaction). 
Let us now introduce a rotation operator $R_{\vec{n}} = e^{-i\alpha S_{\vec{n}}}$, where $S_{\vec{n}}=\sin{\theta}\cos{\phi}S_x+\sin{\theta}\sin{\phi}S_y+\cos{\theta}S_z$ is a collective spin observable along any given direction $\vec{n}=\left( \sin{\theta}\cos{\phi},\sin{\theta}\sin{\phi},\cos{\theta} \right)$. Applying the rotation operator, we can adjust the squeezing axis of OAT interaction
\begin{align}
R^B_{\vec{n}} e^{i \frac{\mu_2}{2} \left( S_z^B \right)^2 } \left( R^B_{\vec{n}} \right)^{\dagger} =  e^{i \frac{\mu_2}{2} \left( S^B_{\vec{n}} \right)^2 }.
\end{align}
By optimizing the twisting axis $\vec{n}$ in the MAI operator $U_{\text{MAI}}(\vec{n})=e^{i\frac{\mu_2}{2} \left( S_{\vec{n}}^B  \right)^2}$, we can obtain a higher steering violation than the one from a fixed twisting axis (the standard z-direction). In Fig.~\ref{FigSM_OATopt}, we compare the steering coefficients with and without optimizing the OAT squeezing axis. It is found that the advantage from  optimizing the direction of the OAT twisting axis is limited, which means performing solely the z-axis OAT interaction is the most efficient and practical choice for experiments.

\begin{figure}[h]
    \begin{center}
	\includegraphics[width=90mm]{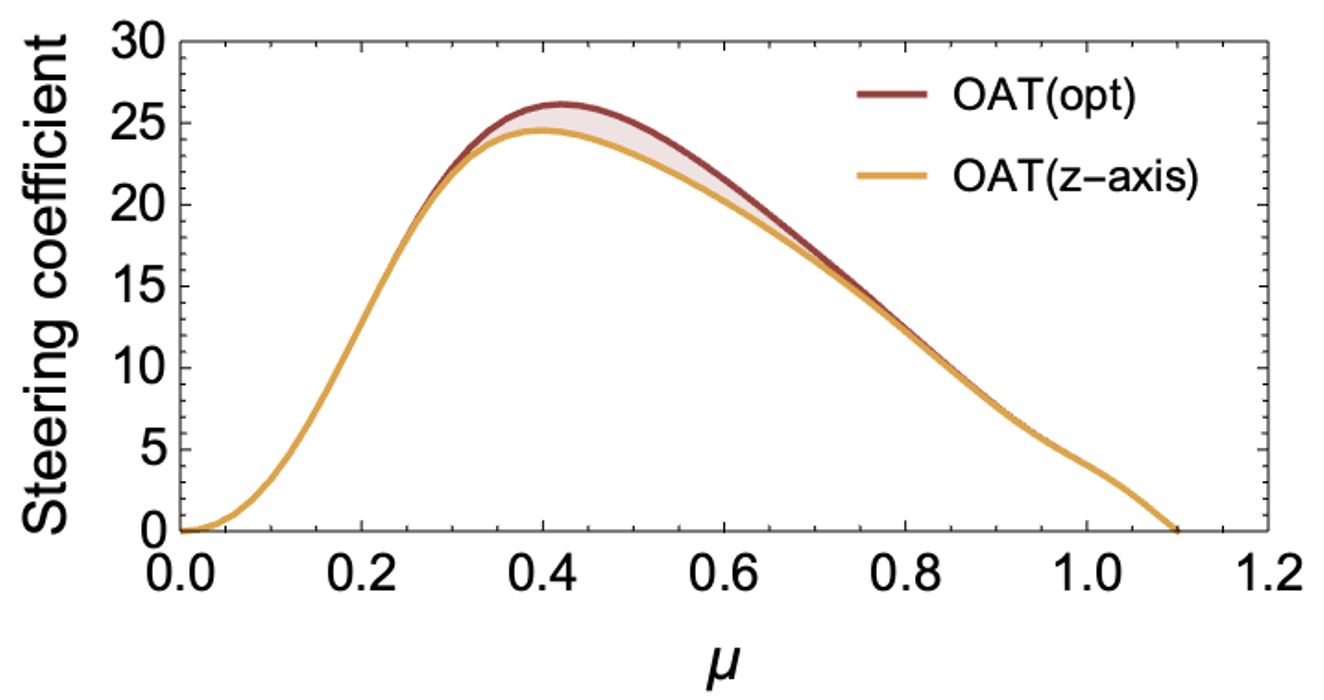}
	\end{center}
    \caption{ For a split spin states with atoms $N=20$, we compare the squeezing coefficients $\delta^R_{\text{MAI}}$ when the MAI unitary operator is $U_{\text{MAI}}=e^{i\frac{\mu_2}{2} \left( S_z^B \right)^2}$ (yellow line) and $U_{\text{MAI}}(\vec{n}_{\text{opt}})=e^{i\frac{\mu_2}{2} ( S_{\vec{n}_{\text{opt}} }^B  )^2}$ with the optimal rotation axis $\vec{n}_{\text{opt}}$ (brown line). } 
    \label{FigSM_OATopt}
\end{figure}

\subsection{C. MAI technique for a general quantum channel}

Although MAI strategies are often based on performing a unitary transformation before measurement, let us remember that in quantum mechanics the most general state evolution is given by a quantum channel. 
For simplicity, we start considering unitary channels, which intuitively are also the one expected to give highest performances, namely channels that can be described as completely positive trace preserving (CPCT) maps 
\begin{equation}
    \rho \rightarrow \Lambda[\rho] = \sum_i K_i \rho K_i^\dagger \;,
\end{equation}
where $K_i$ are Kraus operators such that $\sum_i K^\dagger_i K_i =1$ (the channel is pure). Optimization over all possible Kraus operators is challenging, therefore one usually consider the evolution under some specific quantum channel.

Let us consider a non-unitary MAI scenario. An unknown phase $\theta$ is first encoded into Bob's conditional states, $\rho^B_{b|Y}(G,\theta)=e^{-iG\theta}\rho^B_{b|Y} e^{iG\theta}$, then the probe will go through a quantum channel via $\rho^B_{b|Y}(G,\theta)\rightarrow\Lambda[\rho^B_{b|Y}(G,\theta)]$ and the resulting state is labeled as $\rho^B_{{b|Y},\Lambda}=\Lambda[\rho^B_{b|Y}(G,\theta)]$. The average squeezing parameters can be written as
\begin{align}
( \chi^{-2} )^{B|A} [\mathcal{A},G,M,Y,\Lambda] &:= \frac{
 \left|\partial_{\theta} \langle M \rangle_{\rho^B_{\Lambda}} \right|^2
 }{ \sum_b p(b|Y) \text{Var}[\rho^B_{{b|Y},\Lambda} ,M] } \Big|_{\theta=0}.
\end{align}
Here $M\in \mathcal{L}$ solely represents the final linear measurements, independent on the MAI evolution. $\rho^B_{\Lambda}=\sum_b p(b|Y) \rho^B_{{b|Y},\Lambda}$ is Bob's reduced states. Thus, the maximum violation of Reid's criterion is 
\begin{align}
\delta^R_{\text{MAI}} &= \max_{X,Y,G,M\in \text{span}(\mathcal{L}),\Lambda} \left( ( \chi^{-2} )^{B|A} [\mathcal{A},G,M,Y,\Lambda] - 4\text{Var}^{B|A}[\mathcal{A},G,X] \right). 
\end{align}

To explore the MAI technique under non-unitary quantum channels, we analyze the relevant example of atom loss in split spin squeezed states, which is a common source of noise in atomic experiments. Bob's conditional states under such loss channel can be expressed by the master equation
\begin{align}
\dot{\rho}^B_{b|Y}(G,\theta) = - i[H, \rho^B_{b|Y}(G,\theta)] + \sum_{i=1}^2 \left( L_i \rho^B_{b|Y}(G,\theta) L_i^\dagger-\frac{1}{2} \{ L_i^\dagger L_i, \rho^B_{b|Y}(G,\theta) \} \right),
\end{align}
where $H=e^{i\chi t_2 S^2_z}$ is the OAT interaction, and $L_1=\sqrt{\gamma} a^B, L_2=\sqrt{\gamma} b^B$ represent the jump operators for one-body loss on Bob's two internal spin states, with equivalent loss rate $\gamma$. In Fig.~\ref{FigSM_Loss}, we show the effects of the atom loss on the steering coefficient $\delta^R_{\text{MAI}}$. Although $\delta^R_{\text{MAI}}$ gradually decreases with increase of the loss rate, it can still hold its advantage over the linear measurements strategy within a certain range of losses. This result shows the robustness of our protocol in the presence of losses.

\begin{figure}[h]
    \begin{center}
	\includegraphics[width=\textwidth]{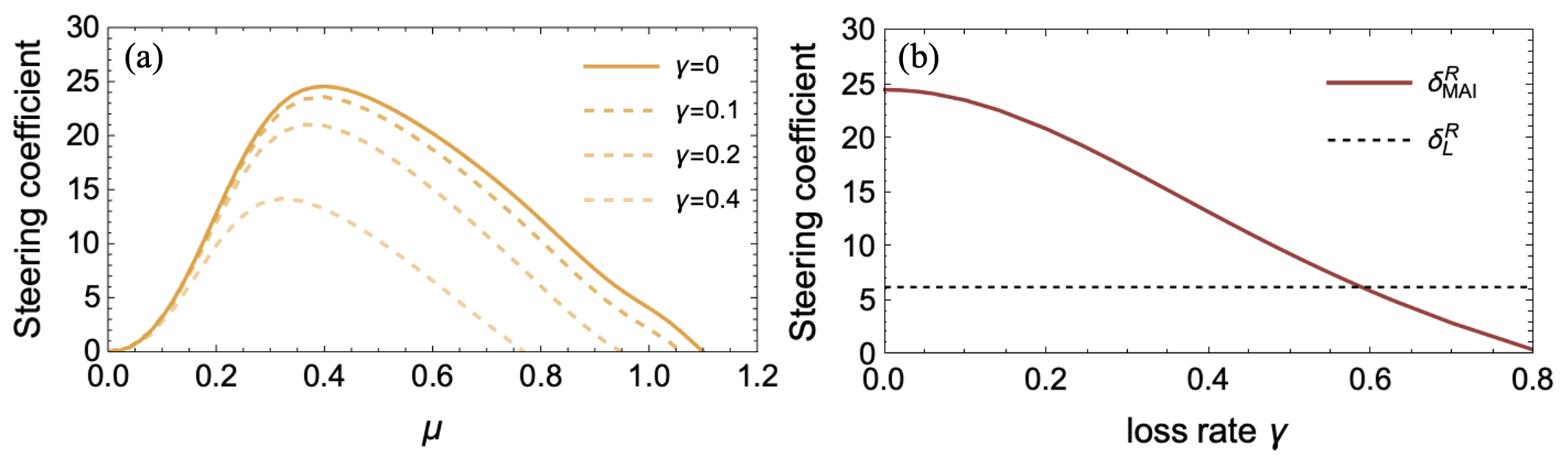}
	\end{center}
    \caption{  The effects of atom loss on the MAI technique for a split spin states with atoms $N=20$. We consider the one-body loss on Bob's two states as $\sqrt{\gamma} a^B, \sqrt{\gamma} b^B$. (a) The steering coefficients $\delta_{\text{MAI}}^R$ change as a function of time evolution $\mu$ under different loss rates $\gamma=\{0,0.1,0.2,0.4\}$. (b) At $\mu=0.4$, $\delta_{\text{MAI}}^R$ (red line) changes with the loss rate $\gamma$. We also plot $\delta_{\text{L}}^R$ (black dashed line) at the same time $\mu=0.4$ for comparison. } 
    \label{FigSM_Loss}
\end{figure}

\subsection{D. Linear-estimate Reid's criterion}
If Bob's estimator linearly depends on Alice's measurement result, the inference variances become $\text{Var}[G_{\text{est}},X]= \text{Var}[G+g_X X]$ and $\text{Var}[M_{\text{est}},Y]= \text{Var}[M+g_Y Y]$, which yield an extensively used linear-estimate Reid's criterion
\begin{align} \label{eq:LEReid}
\Delta^R = \frac{ |\langle [G,M] \rangle_{\rho^B}|^2 }{ \text{Var}[M+g_Y Y] } -4\text{Var}[G+g_X X] \leq 0,
\end{align}
where $g_X,g_Y$ are two real constants that minimize the inference variances. For Gaussian states, the optimized linear-estimator is the best estimator, i.e. $\text{Var}[M+g_Y Y]=\text{Var}[\mathcal{A},M,Y]$, and $\Delta^R$ in Eq.~\eqref{eq:LEReid} is a necessary and sufficient criterion for detecting EPR steering~\cite{ReidRMP2009}.

We have detected steering for a two-mode squeezed state, which is a typical case of continuous-variable (CV) Gaussian states. The state can be prepared by performing a two-mode squeezing operator $S_{AB}=\text{exp}\big[-\xi a^\dagger b^\dagger+\xi^* a b \big]$ with $\xi=re^{i\theta}$ on a vacuum states $|\psi\rangle =S_{AB}|0,0\rangle$. Without loss of generality, we can set $\theta=0$. A complete family of the first-order moments is constructed by two quadrature observables $\mathcal{L}=\{x,p\}$. Considering $M,X$ as local quadrature operators $x^\alpha$ on mode $\alpha=\{A,B\}$, and $G,Y$ as local quadrature operators $p^\alpha$, we can yield the maximum violation of the linear-estimate Reid's criterion in the typical scenario,
\begin{align}\label{eq:leRlinear}
\delta^R_{\text{L}} &=\max_{g_X,g_Y} \frac{1}{ \text{Var} [x^B + g_Y x^A]} -4 \text{Var} [p^B+g_X p^A] \nonumber \\
&= 2\cosh{(2r)} - \frac{2}{ \cosh{(2r)}}.
\end{align}
Here, the optimized constants are $g_{Y,\text{opt}}=-g_{X,\text{opt}}=\tanh{(2r)}$. 

Then we investigate the MAI protocol, where an additional evolution on Bob's side is given by the one-mode squeezing operator $S_B=\text{exp}\big[ r_2  (b^{\dagger 2}-b^2)/2 \big]$. The MAI operator can be expressed as $M_{\text{MAI}}=S_B^\dagger M_{\text{L}} S_B$. Therefore, the maximum violation of Reid's criterion in the MAI scenario is
\begin{align}\label{eq:leRMAI}
\delta^R_{\text{MAI}} &= \max_{g_X,g_Y} \frac{ | \langle [ p^B, S_B^\dagger x^B S_B ] \rangle_{\rho^B} |^2  }{ \text{Var}[ S_B^\dagger  x^BS_B  +g_X x^A ] } -4\text{Var}[p^B+g_Y p^A] \nonumber \\
&= 2\cosh{(2r)} - \frac{2}{ \cosh{(2r)}},
\end{align}
where the optimized constants are $g_{Y,\text{opt}}=e^{r_2}\tanh{(2r)}, g_{X,\text{opt}}=-\tanh{(2r)}$. It is found that the sensitivity in Eq.~\eqref{eq:leRMAI} is independent with the MAI squeezing $r_2$, and it is equivalent with the sensitivity from the typical protocol in Eq.~\eqref{eq:leRlinear}. 

When detection noise is present, the measurements can be reexpressed as $\tilde{x}^\alpha=x^\alpha+\Delta x^\alpha$ and $\tilde{p}^\alpha=p^\alpha+\Delta p^\alpha$, where $\Delta x^\alpha, \Delta p^\alpha$ are random variables following a Gaussian distribution with mean $\langle \Delta M \rangle =0$ and variance $\langle (\Delta M )^2\rangle=\sigma^2$. Under the influence of detection noises, the maximum violation in a typical scenario reads
\begin{align}
\delta^R_{\text{L}} = \frac{1}{ \frac{1}{2\cosh{(2r)}} +(1+\tanh^2{(2r)}) \sigma^2 } - 4 \left( \frac{1}{2\cosh{(2r)}}+(1+\tanh^2{(2r)}) \sigma^2 \right).
\end{align}
In the MAI protocol, the MAI operator with detection noise can be reexpressed as $S_B^\dagger \tilde{x}^B S_B =S_B^\dagger  x^B S_B +\Delta x^B$, which leads to the maximum violation
\begin{align}
\delta^R_{\text{MAI}} = \frac{ e^{2r_2} }{  \frac{e^{2r_2}}{ 2\cosh{(2r)} } +(1+e^{2r_2} \tanh^2{(2r)}) \sigma^2 } - 4 \left( \frac{1}{2\cosh{(2r)}}+(1+\tanh^2{(2r)}) \sigma^2 \right).
\end{align}
Therefore, we find that the criterion with MAI technique is more robust to detection noise than the typical one as long as $r_2>0$. The violation value $\delta^R_{\text{MAI}}$ increases with $r_2$ and reaches its maximum when $r_2$ is infinity, which reads
\begin{align}
\delta^R_{\text{MAI}} (r_2\rightarrow \infty)= \frac{1+\cosh{(4r)}}{ \cosh{(2r)}+2\sigma^2 \sinh^2{(2r)} } - 4 \left( \frac{1}{2\cosh{(2r)}}+(1+\tanh^2{(2r)}) \sigma^2 \right).
\end{align}

\section{II. \quad Steering criterion based on quantum Fisher information}
Besides considering the complementarity between two non-commuting observables as in Reid's criterion, the work in \cite{MatteoNC2021} formulated a shaper steering criterion based on the complementarity principle between a phase and its generator. This criterion reads
\begin{align}
F_Q^{B|A} [\mathcal{A},G]-4\text{Var}_Q^{B|A} [\mathcal{A},G] \leq 0,
\end{align}
which is determined by the quantum conditional Fisher information
\begin{align}
    F_Q^{B|A}[\mathcal{A},G]:= \max_Y \sum_b p(b|Y) F_Q[\rho^B_{b|Y},G]
\end{align}
and the quantum conditional variance 
\begin{align}
\text{Var}_Q^{B|A} [\mathcal{A},G]:= \min_{X} \sum_a p(a|X) \text{Var}[\rho_{a|X}^B,G].
\end{align}
For simplicity, if we consider the specific measurements $X,Y$ for Alice rather than optimizing over them, we will obtain a weaker criterion as Eq.~\eqref{eq:DeltaF} in the main text,
\begin{align}\label{eq:SMDeltaF}
\Delta^F := F^{B|A}[\mathcal{A},G,Y] - 4\text{Var}[\mathcal{A},G,X] \leq 0.
\end{align}
The relation $F^{B|A}[\mathcal{A},G,Y] \geq (\chi^{-2})^{B|A} [\mathcal{A},G,M,Y]$ leads to the hierarchy between two criteria $\Delta^F \geq \Delta^R$.

Because the split spin states are pure states, Bob's conditional states $\rho^B_{b|Y}$ are pure as well. For any pure states, the quantum Fisher information can be expressed by the associated variance, so that we have $F_Q[\rho^B_{b|Y},G]=4\text{Var}[\rho^B_{b|Y},G]$. In this case, the conditional Fisher information can be written as 
\begin{align}
F^{B|A}[\mathcal{A},G,Y] &:= \sum_b p(b|Y) F_Q [\rho^B_{b|Y},G] \nonumber \\
&= 4 \sum_b p(b|Y) \text{Var} [\rho^B_{b|Y},G] \nonumber \\
&= 4 \text{Var}^{B|A}[\mathcal{A},G,Y].
\end{align}
Using the approach in Eq.~\eqref{eq:Reid2ndterm}, we can optimize the criterion in Eq.~\eqref{eq:SMDeltaF} over the measurements as 
\begin{align}
\max_{G\in \text{span}(\mathbf{G})} \Delta^F &= \max_{G\in \text{span}(\mathbf{G})} \left( 4 \text{Var}^{B|A}[\mathcal{A},G,Y] - 4 \text{Var}^{B|A}[\mathcal{A},G,X] \right) \nonumber \\
&= 4 \max_{\mathbf{n}} \left( \mathbf{n}^T \boldsymbol{\Gamma}^{B|A}[\mathcal{A},G,Y] \mathbf{n} -  \mathbf{n}^T\boldsymbol{\Gamma}^{B|A}[\mathcal{A},G,X] \mathbf{n} \right) \nonumber \\
&= 4 \lambda_{\max} \left( \boldsymbol{\Gamma}^{B|A}[\mathcal{A},G,Y]  -  \boldsymbol{\Gamma}^{B|A}[\mathcal{A},G,X] \right).
\end{align}
Restricting Alice's measurements $X,Y$ to $yz-$plane, we will finally obtain the maximum violation 
\begin{align}
\delta^F := 4 \max_{\theta_X,\theta_Y} \lambda_{\max} \left( \boldsymbol{\Gamma}^{B|A}[\mathcal{A},G,Y]  -  \boldsymbol{\Gamma}^{B|A}[\mathcal{A},G,X] \right),
\end{align}
where the eigenvector associated to the maximum eigenvalue represents the optimal measurement direction $\mathbf{n}_{\text{opt}}$.

\section{III. Giovannetti's entanglement criterion}
Besides detecting EPR steering, we also investigate on detecting entanglement with MAI technique. We use Giovannetti's criterion to detect the entanglement between the two ensembles in split spin squeezed states, which reads
\begin{align}
\Delta^G := \frac{ ||g_X g_Y| \langle[ X,Y ] \rangle + \langle [G,M] \rangle |^2 }{ \text{Var}[M+g_Y Y] } -4\text{Var}[G+g_X X] \leq 0.
\end{align}
Here $g_X,g_Y$ are real constants chosen to optimize the criterion. When we consider the measurement directions for $M,Y$ along the squeezing direction $z'$, while the measurement directions for $G,X$ along the anti-squeezing direction $y'$, the violation for Giovannetti's criterion is given as
\begin{align}
\delta^G_{\text{L}} = \max_{g_X,g_Y} \frac{ | \langle S_x^A \rangle + \langle S_x^B \rangle  |^2  }{ \text{Var} [S_{z'}^B+ g_Y S_{z'}^A] } -4\text{Var}[S_{y'}^B +g_X S_{y'}^A] .
\end{align}
Here, $S_{z'}^\alpha = -\sin{\theta_{S}} S_y^\alpha+\cos{\theta_S} S_z^\alpha$ and $S_{y'}^\alpha = \cos{\theta_{S}} S_y^\alpha+\sin{\theta_S} S_z^\alpha$ are determined by the squeezing angle 
\begin{align}
\theta_S =\frac{1}{2} \arctan{\left( \frac{4 \sin{\left( \frac{\mu}{2}\right)}  \cos^{N-2}{\left( \frac{\mu}{2} \right)} }{ 1-\cos^{N-2}{(\mu)} } \right)}.
\end{align}
The Giovannetti's criterion with only linear collective spin observables involved has also been investigated in \cite{YumangNJP2019,MatteoScience2018}.

In the scenario with MAI technique, an additional unitary evolution $U(\mu_2)=e^{i\mu_2/2 (S_z^B)^2}$ is involved, and the criterion becomes
\begin{align}
\Delta^G_{\text{MAI}}=  \frac{ \big| |g_X g_Y| \langle[ X,Y ] \rangle + \langle [G,U^\dagger (\mu_2) M_L U(\mu_2)] \rangle \big|^2 }{ \text{Var}[ U^\dagger (\mu_2) M_L U(\mu_2)+g_Y Y] } -4\text{Var}[G+g_X X] \leq 0.
\end{align}
If we restrict linear collective spin observables to $yz-$plane, the maximum violation can be obtained after optimization over measurements and constants
\begin{align}
\delta^G_{\text{MAI}} = \max_{g_X,g_Y,\mu_2} \max_{X,Y,H,M_L \in \text{span}(\mathcal{L}_{yz})} \Delta^G_{\text{MAI}},
\end{align}
where $\mathcal{L}_{yz}=\{S_y,S_z\}$.

\end{widetext}

\end{document}